\documentclass[letterpaper,journal]{IEEEtran}
\usepackage{amsmath,amsfonts}
\usepackage{array}
\usepackage[caption=false,font=footnotesize,labelfont=footnotesize,textfont=normalsize]{subfig}
\usepackage{textcomp}
\usepackage{stfloats}
\usepackage{url}
\usepackage{verbatim}
\usepackage{graphicx}
\usepackage{cite}
\usepackage{setspace} 
\usepackage[linesnumbered,ruled]{algorithm2e}
\usepackage{multirow}
\usepackage{amssymb,bm}

\newcommand{\inc}{\text{inc}}
\newcommand{\tot}{\text{tot}}

\graphicspath{{Figures/}}
\usepackage[colorlinks,
            linkcolor=blue,
            anchorcolor=blue,
            citecolor=blue]{hyperref}

\begin{document}

\title{Electromagnetic Quantitative Inversion for Translationally Moving Targets via Phase Correlation Registration of Back-Projection Images}

\author{Yitao~Lin, Dahai Dai, Shilong Sun, Yuchen Wu, and Bo Pang
\thanks{This work was supported by the National Natural Science Foundation of China under Grant 62471476 and Grant 62001485. \textit{(Corresponding author: Shilong Sun)}}
\thanks{Yitao~Lin, Dahai Dai, Shilong Sun, Yuchen Wu, and Bo Pang are with the College of Electronic Science and Technology, National University of Defense Technology, Changsha 410073, China (e-mail: sunshilong@nudt.edu.cn).}}

\markboth{IEEE antennas and wireless propagation letters,~Vol.~xx, No.~x, xx~2025}%
{Shell \MakeLowercase{\textit{et al.}}: A Sample Article Using IEEEtran.cls for IEEE Journals}

\maketitle

\begin{abstract}
A novel electromagnetic quantitative inversion scheme for translationally moving targets via phase correlation registration of back-projection (BP) images is proposed. Based on a time division multiplexing multiple-input multiple-output (TDM-MIMO) radar architecture, the scheme first achieves high-precision relative positioning of the target, then applies relative motion compensation to perform iterative inversion on multi-cycle MIMO measurement data, thereby reconstructing the target’s electromagnetic parameters. As a general framework compatible with other mainstream inversion algorithms, we exemplify our approach by incorporating the classical cross-correlated contrast source inversion (CC-CSI) into iterative optimization step of the scheme, resulting in a new algorithm termed RMC-CC-CSI. Numerical and experimental results demonstrate that RMC-CC-CSI offers accelerated convergence, enhanced reconstruction fidelity, and improved noise immunity over conventional CC-CSI for stationary targets despite increased computational cost. 
\end{abstract}


\section{Introduction}\label{sec.introduction}
\IEEEPARstart{E}{lectromagnetic} (EM) inverse scattering imaging restores the electromagnetic scattering echo received by antennas to the permittivity or conductivity distribution image of target object. It has been widely used in medical imaging, through wall imaging, oil reservoir exploration and subsurface object detection \cite{2021Machine,2023Real-Time,Chu2019Fast,2020A,Hajebi2018An,10839289,10787043}, etc. EM inverse scattering problems (ISPs) are inherently ill-posedness and nonlinear. Traditional quantitative iterative inversion methods like Born iterative method (BIM) \cite{Y.M.Wang1989An}, contrast source inversion (CSI) \cite{van1997contrast,2001Contrast}, subspace-based optimization method (SOM) \cite{XudongChen2010} and CC-CSI \cite{sun2017Cross,Sun2018Inversion} share better performance and stronger applicability than non-iterative methods, but with larger computational cost. 

Despite remarkable advances in EM ISPs for static targets, EM quantitative imaging of moving targets remains underdeveloped. Current techniques for reconstructing dielectric properties and velocity profiles \cite{Pastorino1,Pastorino2,Praveen} are constrained by simplistic assumptions on target geometry, material composition, and motion patterns. Such approaches fail to handle geometrically irregularity or complex motions encountered in practice. Moreover, they only provide a global dielectric estimate, falling short of quantitatively resolving permittivity and conductivity distributions for complex components. Inspired by ISAR \cite{1980Principles,2014Inverse,2010An}, which exploits relative target-radar motion for high azimuth resolution and has a natural advantage in imaging of moving targets, this paper investigates the technical feasibility of EM quantitative inversion imaging for moving targets, explores the potential of EM inversion techniques.  

We apply TDM-MIMO architecture to sequentially activate transmitters with identical waveforms, collecting target echoes in non-overlapping time slots, which eliminates inter-signal interference, simplifies signal separation. For brevity, it is abbreviated as MIMO architecture. When a moving target enters the observation coverage of a MIMO system, we first propose two critical simplifying assumptions to systematically advance our research:
\begin{enumerate}
\item \label{aspt.1} Given the mature high scanning rate of modern MIMO systems, we assume the target’s velocity is substantially lower than the array’s scanning rate. This allows us to approximately neglect the influence of Doppler effects and adopt an analogy to the stop-go-stop model where the target remains stationary during each full MIMO scanning cycle. Thus, each cycle's measurement data corresponds to a distinct instantaneous spatial target configuration.
\item \label{aspt.2} The target is modeled as an ideal rigid body undergoing purely translational motion within the observation coverage, with no rotational orientation changes.
\end{enumerate}
Building upon above assumptions, we first obtain the simplified translationally target's spatial positions and distributions at different instants by performing radar BP \cite{161068BP} imaging of moving targets across multiple MIMO observation cycles. A phase correlation image registration method \cite{Kuglin1975The,Zhou2010Phase} is then conducted to estimate relative displacements between these instances using one instance as a reference. By converting displacements into equivalent antenna position corrections, we integrate relative motion compensation into the classical CC-CSI framework, forming a new method termed RMC-CC-CSI that achieves quantitative inversion for moving targets. Evaluated with synthetic data, RMC-CC-CSI demonstrates accelerated convergence, superior reconstruction accuracy, and enhanced noise immunity over conventional CC-CSI for stationary targets. Experimental results validate its effectiveness. 
\section{Problem Statement}\label{sec.Formulation}
\begin{figure}[!t]
    \centering
    \includegraphics[width=0.55\linewidth]{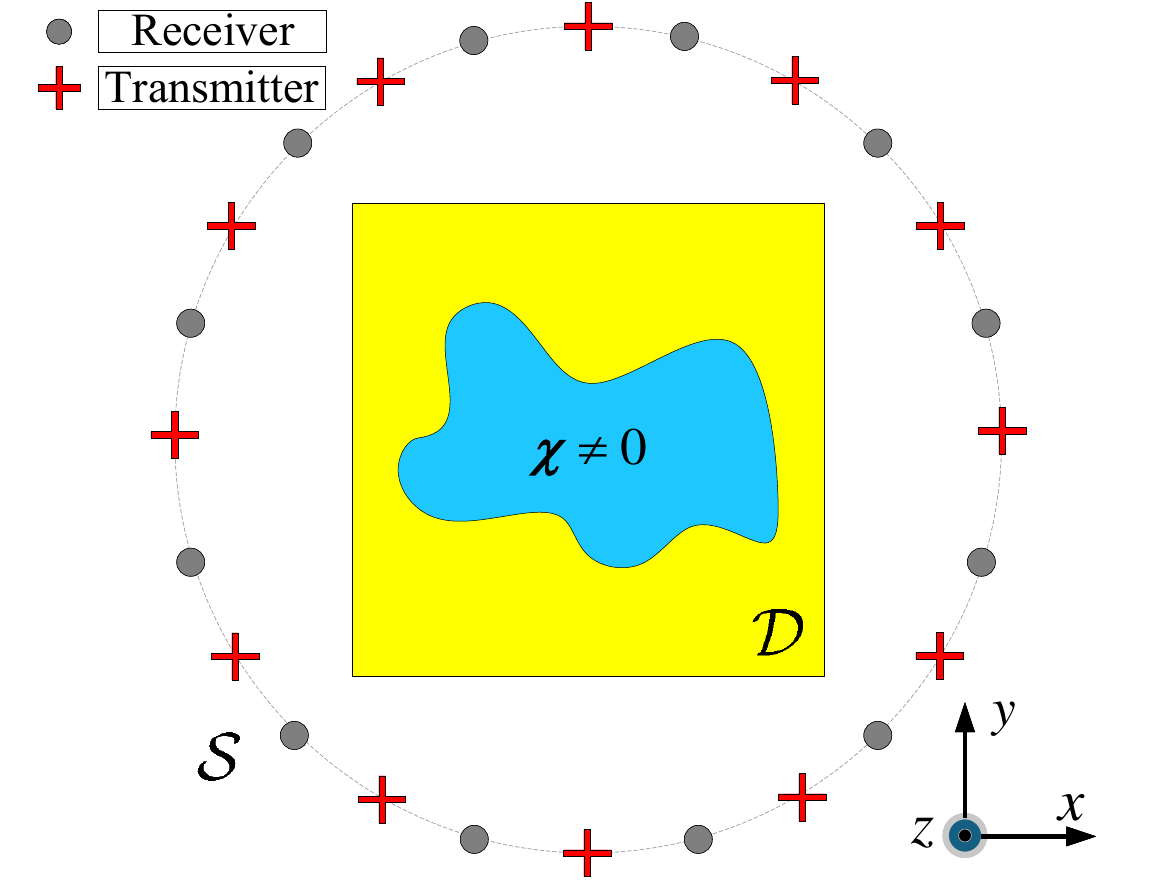}%
    \caption{Configuration of the 2-D EM ISP under TM polarization.}
    \label{fig:ProblemStatement}   
\end{figure}
In the free space background, the 2-D EM ISP configuration under transverse magnetic (TM) incidence is shown in Fig.~\ref{fig:ProblemStatement}. An unknown scatterer with contrast $\bm{\chi}$ lies within the imaging domain $\mathcal{D}$, illuminated by TM waves with time factor of $e^{\text{i} \omega t}$, where $\text{i}=\sqrt{-1}$ represents the imaginary unit and $\omega$ is the angular frequency. The measurement domain $\mathcal{S}$ contains transmitters denoted by $p\in\{1,2,3...,P\}$ and receivers denoted by $q\in\{1,2,3,...,Q\}$ that record the scattered field. We use bold symbols for vectors with three components. The data equation and state equation based on the finite-difference frequency-domain (FDFD) \cite{W.Shin2013} are:
\begin{equation}
    {\bm{f}_{p}}={{\mathcal{M}}^{\mathcal{S}}}{\bm{A}^{-1}}{{\omega }^{2}}\bm{\chi} \odot \bm{e}_{p}^\tot,\ \ {\bm{f}_{p}}\in \mathcal{S}
\label{dataEq}
\end{equation}
\begin{equation}
    \bm{e}_{p}^{\tot}=\bm{e}_{p}^{\inc}+{{\mathcal{M}}^{\mathcal{D}}}{\bm{A}^{-1}}{{\omega }^{2}}\bm{\chi} \odot \bm{e}_{p}^{\tot},\ \ \bm{e}_{p}^{\tot}\in \mathcal{D}
\label{stateEq}
\end{equation}
where $\bm{f}_p$ is the measurement data of the scattered field, $\bm{A}$ is the stiffness matrix under FDFD scheme. $\mathcal{M}^{\mathcal{S}}$ is an operator that interpolates the field values defined on the finite difference grid to the receiver position, $\mathcal{M}^{\mathcal{D}}$ is an operator that obtains the field values in the field domain $\mathcal{D}$, which is uniformly discretized into $N$ square grids of equal area. $\bm{e}^\tot_{p}$ and $\bm{e}^\inc_{p}$ represent the electric total field and incident field, the contrast $\bm{\chi}=\bm{\varepsilon}-\bm{\varepsilon}_{bg}$, where $\bm{\varepsilon} = \bm{\epsilon} - \text{i}\bm{\sigma}/{\omega}$ and $\bm{\varepsilon}_{bg} = \bm{\epsilon}_{bg} - \text{i}\bm{\sigma}_{bg}/{\omega}$ are the complex permittivity distributions with and without targets. $\bm{j}_{p}=\bm{\chi} \odot \bm{e}^\tot_p$ is defined as the contrast source, where $\odot$ is the component multiplication operator. For the sake of brevity, $\omega^2$ will be written into $\bm{A}$ in the rest of this paper. The ISP is to reconstruct the contrast $\bm{\chi}$ through the incomplete scattered field measurement data $\bm{f}_p$.
\section{Translationally Moving Target Inversion}\label{sec.Method}
The proposed scheme consists of two components: fast positioning based on phase correlation registration of BP images, followed by iterative inversion for moving targets (RMC-CC-CSI). The related flowchart is shown in Fig.~\ref{fig:RMC_CCCSI_Flow}.
\subsection{Phase Correlation Registration of BP Images}The radar system captures MIMO data in each scanning cycle during moving target observation. By applying BP imaging, which has been widely adopted in radar imaging applications \cite{Wang2010MIMO}, to data from multiple cycles, we obtain a series of target 'snapshots' that reveal its positional information at each instant. For 2-D scenarios, let ${\bm{x}} = [x_1, x_2, \ldots, x_{N_\text{M}}]$ and ${\bm{y}} = [y_1, y_2, \ldots, y_{N_\text{M}}]$ represent the target's ideal center coordinates across $N_\text{M}$ observation instants, where $N_{\text{M}}$ being the total number of MIMO scanning instances. Preliminary BP imaging localizes the target coarsely, but perspective variations and target extent cause shape inconsistencies between successive images. These deformations impede reliable motion centroid determination and precise spatiotemporal localization. We introduced the phase correlation algorithm \cite{Kuglin1975The,Gui2004Phase} for image registration to obtain translational shifts between consecutive frames. Selecting a central reference BP imaging result at the $c$-th observation instant (where $c \in \{1,2,\ldots,N_{\text{M}}\}$) to confirm the basic inversion domain $\mathcal{D}^{c}$ of the moving target, then the BP results are used for registration to obtain $({\bm{x}_{r,{c}}},{\bm{y}_{r,{c}}})$, the spatial relative displacement at different instants referencing to the position at specific instant $c$, which satisfies:
\begin{figure}[!t]
    \centering
    \includegraphics[width=1\linewidth]{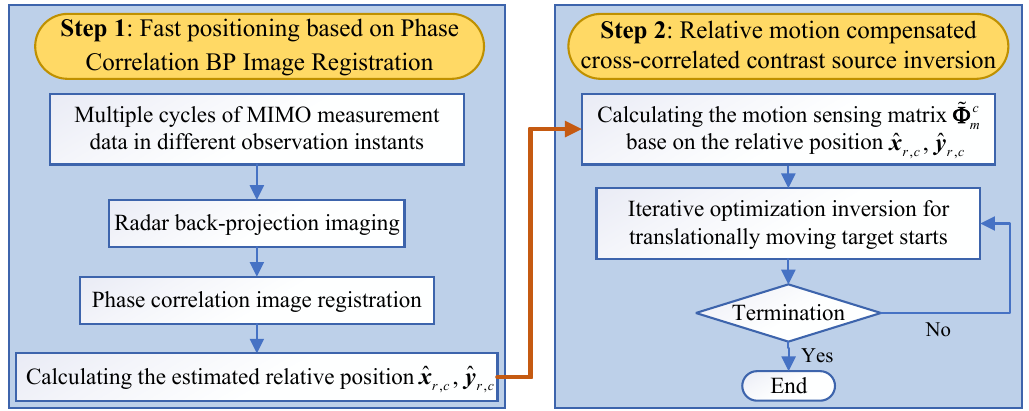}%
    \caption{Flowchart of the proposed scheme.}
    \label{fig:RMC_CCCSI_Flow}
\end{figure}
\begin{equation}
    ({\bm{x}_{r,{c}}},{\bm{y}_{r,{c}}}) = ({\bm{x}} - x_{c} ,{\bm{y}} - y_{c}) 
\label{eq.RelaPos}
\end{equation}

\subsection{Relative Motion Compensated CC-CSI Based on FDFD} 
During target motion, all elements defined over the discretized imaging domain $\mathcal{D}$ including the sensing matrix $\bm{\Phi}=\mathcal{M}^{\mathcal{S}}\bm{A}^{-1}$, contrast source $\bm{j}_p$ and contrast $\bm{\chi}$ distributions, become coupled with the target's dynamic positions. Based on \cite{Sun2017Qualitative}, we construct the motion sensing matrix : $\tilde{\bm{\Phi}} _{m}^{c} = \mathcal{M}_{m}^{\tilde{\mathcal{S}}^c} \bm{A}^{-1}$, where $\mathcal{M}_{m}^{\tilde{\mathcal{S}}^c}$ denotes the motion-compensated $\mathcal{M}^{\mathcal{S}}$ operator referenced to central instant $c$, and $m \in \{1,2,\ldots,N_{\text{M}}\}$ indexes different observation instants. For TM polarization in homogeneous media, ${\tilde{\bm{\Phi}}_{m}^{c}}$ is analytically formulated as:
    \begin{equation}
        {\tilde{\bm{\Phi}}_{m}^{c}}[q,n]=\frac{\text{i}{{\omega }}{\mu }_{0}}{4} H_{0}^{(1)}(-k{{\left\| {\bm{r}_{m,q}^{c}}-{\bm{r}_{n}^{c}} \right\|}_{2}})
    \label{eq.MotionSensingMatrix}
    \end{equation}
$\tilde{\bm{\Phi}}_{m}^{c}[q,n]$ is the element at row $q$ and column $n$ of the matrix ${\tilde{\bm{\Phi}}_{m}^{c}}$; $H_0^{(1)}(\cdot)$ represents the Hankel function of first kind with order zero; $k$ is the wavenumber; ${\bm{r}_{m,q}^{c}}$ denotes the relative position vector of the $q$-th receiver in the $m$-th MIMO scanning round, referenced to the target's position at the $c$-th observation instant. $\bm{r}_{n}^{c}$ is the $n$-th position vector in the selected reference inversion domain $\mathcal{D}^{c}$. Through phase correlation registration of BP images, the relative position variations between target and transceivers across different observation instants can be obtained and transformed into equivalent relative transceivers' positions vector ${\bm{r}_{m,q}^{c}}$:
\begin{equation}
{\bm{r}_{m,q}^{c}} - {\bm{r}_{c,q}^{c}}= -[{\bm{x}_{r,{c}}}(m),{\bm{y}_{r,{c}}}(m)]
\label{eq.r}
\end{equation}
here ${\bm{r}_{c,q}^{c}}$ is the known position vector of the q-th receiver in the $c$-th MIMO scanning cycle. Accordingly, ${\bm{r}_{m,q}^{c}}$ is used to calculate the motion sensing matrix $\tilde{\bm{\Phi}}_{m}^{c}$ then. With the ignorance of the $\mathcal{M}^{\mathcal{D}}$ operator always coupled with $\bm{A}$, the data error, state error and cross-correlated error suitable for ISP of moving targets are defined as:
\begin{equation}
    {{\tilde{\bm{\rho}}}_{m,p}^{c}}={{\bm{f}}_{m,p}}-{{\tilde{\bm{\Phi}}}_{m}^{c}}{{\bm{j}}_{m,p}^{c}}
\label{MovingDataError}
\end{equation}
\begin{equation}
    {{\tilde{\bm{\gamma}}}_{m,p}^{c}}={{\bm{\chi }}}\odot \bm{e}_{m,p}^{\text{inc},c}+{{\bm{\chi }}}\odot {{\bm{A}}}^{-1}{{\bm{j}}_{m,p}^{c}}-{{\bm{j}}_{m,p}^{c}}
\label{MovingStateError}
\end{equation}
\begin{equation}
    {{\tilde{\bm{\xi}}}_{m,p}^{c}}={{\bm{f}}_{m,p}}-{{\tilde{\bm{\Phi}}}_{m}^{c}}({{\bm{\chi }}}\odot \bm{e}_{m,p}^{\text{inc},c}+{{\bm{\chi }}}\odot {{\bm{A}}}^{-1}{{\bm{j}}_{m,p}^{c}})
\label{MovingCCError}
\end{equation}
    \begin{algorithm}[!t]
        \setstretch{1.0}
        \caption{RMC-CC-CSI}\label{RMC-CC-CSI}
        \SetKwInOut{Input}{Input}
        \SetKwInOut{Output}{Output}
        \Input{$c$, ${{\tilde{\bm{\Phi}}}_{m}^{c}}$, ${\bm{f}}_{m,p}$, ${\bm{j}}_{m,p}^{c}$, $\bm{e}_{m,p}^{\text{inc},c}$}
        Initialize the contrast sources (iteration
counter $\ell$ = 0): ${\bm{j}}_{m,p,0}^{c}=\frac{\left\|(\tilde{\bm{\Phi}}_{m}^{c})^H\bm{f}_{m,p}\right\|^2_{\mathcal{D}^{c}}}{\left\|\tilde{\bm{\Phi}}_{m}^{c}(\tilde{\bm{\Phi}}_{m}^{c})^H\bm{f}_{m,p}\right\|^2_{\tilde{\mathcal{S}}^c}}(\tilde{\bm{\Phi}}_{m}^{c})^H\bm{f}_{m,p}$\;
        Initialize total field $\bm{e}^{\tot,c}_{m,p,0} = \bm{e}^{\inc,c}_{m,p}+ \bm{A}^{-1}\bm{j}_{m,p,0}^c$\;
        Initialize $\tilde{\eta}^{\tilde{\mathcal{S}}^c}$, $\bm{\nu}_{m,p}^c=0$, $\bm{\nu}^{\bm{\chi}}=0$\;
        Initialize contrast $\bm{\chi}=\frac{\sum_{m=1}^{{N}_{\text{M}}}\sum_{p=1}^{P}\bm{j}_{m,p,0}^c\overline{\bm{e}^{\tot,c}_{m,p,0}}}
        {\sum_{m=1}^{{N}_{\text{M}}}\sum_{p=1}^{P}\bm{e}^{\tot,c}_{m,p,0}\overline{\bm{e}^{\tot,c}_{m,p,0}}}$\;
        Set maximum iterations $\ell_{\max}$ and initialize $\ell=1$\; 
        \While{$\ell \leq N_{\max}$}
         {
            Calculate $\tilde{\eta}^{\mathcal{D}^{c}}$ and relative motion compensated data, state and cross-correlated errors (residuals)\;
            $\bm{g}_{m,p}^{\bm{j}^c,\text{old}}\leftarrow \bm{g}_{m,p}^{\bm{j}^c}$, $\bm{g}_{m,p}^{\bm{j}^c}$ is the gradient of \eqref{eq.cost_j}\;
            $\bm{\nu}_{m,p}^c \leftarrow \bm{g}_{m,p}^{\bm{j}^c}+\frac
                 {
                   \sum\limits_{p'}
                   \left\langle 
                     \bm{g}_{m,p'}^{\bm{j}^c}{,}\bm{g}_{m,p'}^{\bm{j}^c}-\bm{g}_{m,p'}^{\bm{j}^c,\text{old}}
                   \right\rangle
                   _{\mathcal{D}^{c}}}
                 {
                   \sum\limits_{p'}
                     \left\|
                       \bm{g}_{m,p'}^{\bm{j}^c,\text{old}}
                     \right\|^2
                     _{\mathcal{D}^{c}}}
                 \bm{\nu}_{m,p}^{c}$\;
            $\bm{e}_{m,p}^{\nu}\leftarrow \bm{A}^{-1}\bm{\nu}_{m,p}^c$, calculate $\alpha$ by minimizing \eqref{eq.cost_j}\;
            ${\bm{j}}_{m,p}^{c} \leftarrow {\bm{j}}_{m,p}^{c}+ \alpha \bm{\nu}_{m,p}^c $, $\bm{e}^{\tot,c}_{m,p} \leftarrow \bm{e}^{\tot,c}_{m,p}+\alpha \bm{e}_{m,p}^{\nu}$\;
            $\bm{g}^{\bm{\chi}}_\text{old} \leftarrow \bm{g}^{\bm{\chi}}$, $\bm{g}^{\bm{\chi}}$ is the gradient of \eqref{eq.cost_chi}\;
            $\bm{\nu}^{\bm{\chi}} \leftarrow
            \bm{g}^{\bm{\chi}} + \frac
            {
            \left\langle 
             \bm{g}^{\bm{\chi}},\ \bm{g}^{\bm{\chi}}-\bm{g}^{\bm{\chi}}_\text{old}
            \right\rangle_{\mathcal{D}^{c}}}
            {\left\|\bm{g}^{\bm{\chi}}_\text{old}\right\|^2_{\mathcal{D}^{c}}}
            \bm{\nu}^{\bm{\chi}}_\text{old}$ \;
            calculate $\beta $ by minimizing \eqref{eq.cost_chi}, $\bm{\chi} \leftarrow \bm{\chi} + \beta \bm{\nu}^{\bm{\chi}}$ \; 
            Increment iteration counter: $\ell \gets \ell+1$\;
         }
        \Output{$\bm{\chi}$, relative permittivity $\hat{\varepsilon}_\text{r}$ and conductivity $\hat{\sigma}$\;}
    \end{algorithm}
    where ${\bm{f}}_{m,p}$ denotes the scattered field acquired through multiple MIMO scanning cycles, ${\bm{j}}_{m,p}^{c}$ and $\bm{e}_{m,p}^{\text{inc},c}$ are distributed in $\mathcal{D}^{c}$. When $N_{\text{M}}=1$, above formulations reduce to a static target imaging scenario. RMC-CC-CSI algorithm minimizes a cost function comprising the three aforementioned error terms, performing alternating updates of ${\bm{j}}_{m,p}^{c}$ and $\bm{\chi}$ during iterative optimization. The cost function for updating the contrast source ${\bm{j}}_{m,p}^{c}$ in RMC-CC-CSI is defined as:
    \begin{equation}
    \begin{split}
    &\mathcal{C}_{\text{RMC-CCCSI}}^{\bm{j}^c}  ={\tilde{\eta }^{{\tilde{\mathcal{S}}^c}}}\sum_{m=1}^{{{N}_{\text{M}}}}{\sum_{p=1}^{P}{\left\| {{\tilde{\bm{\rho}}_{m,p}^{c}}} \right\|_{{\tilde{\mathcal{S}}^c}}^{2}}} \\
    &+{\tilde{\eta }^{\mathcal{D}^{c}}\sum_{m=1}^{{{N}_{\text{M}}}}{\sum_{p=1}^{P}{\left\| {{\tilde{\bm{\gamma}}}_{m,p}^{c}} \right\|_{\mathcal{D}^{c}}^{2}}}}  
    +{\tilde{\eta }^{{\tilde{\mathcal{S}}^c}}}\sum_{m=1}^{{{N}_{\text{M}}}}{\sum_{p=1}^{P}{\left\| {{\tilde{\bm{\xi}}}_{m,p}^{c}} \right\|_{{\tilde{\mathcal{S}}^c}}^{2}}}
    \end{split}
    \label{eq.cost_j}
    \end{equation}
here ${\left\|\cdot\right\|}_{\tilde{\mathcal{S}}^c}$ and ${\left\|\cdot\right\|}_{\mathcal{D}^{c}}$ represent the 2-norms on the relative measurement space $L^2 (\tilde{\mathcal{S}}^c)$ and the imaging field space $L^2 (\mathcal{D}^{c})$. $\tilde{\eta}^{\tilde{\mathcal{S}}^c}=1/(\sum_{m=1}^{{N}_{\text{M}}}{\sum_{p=1}^{P}{\left\| {{\bm{f}}_{m,p}} \right\|}_{{\tilde{\mathcal{S}}^c}}^{2}})$, $\tilde{\eta}^{\mathcal{D}^{c}} = 1/ (\sum_{m=1}^{{N}_{\text{M}}}{\sum_{p=1}^{P}{\left\| {{\bm{\chi }}}\odot \bm{e}_{m,p}^{\text{inc},c} \right\|}_{\mathcal{D}^{c}}^{2}})$. The cost function for updating $\chi$ is defined as:
    \begin{equation}
     \mathcal{C}_{\text{RMC-CCCSI}}^{\bm{\chi}} ={\tilde{\eta }^{\mathcal{D}^{c}}\sum_{m=1}^{{{N}_{\text{M}}}}{\sum_{p=1}^{P}{\left\| {{{\tilde{\bm{\gamma}}}}_{m,p}^{c}} \right\|_{\mathcal{D}^{c}}^{2}}}}
     +{\tilde{\eta }^{{\tilde{\mathcal{S}}^c}}}\sum_{m=1}^{{{N}_{\text{M}}}}{\sum_{p=1}^{P}{\left\| {{{\tilde{\bm{\xi}}}}_{m,p}^{c}} \right\|_{{\tilde{\mathcal{S}}^c}}^{2}}}
    \label{eq.cost_chi}
    \end{equation}

    Detailed implementation for RMC-CC-CSI is given in Algorithm~\ref{RMC-CC-CSI}, maintaining identical algorithmic procedures to standard CC-CSI. Here, $\overline{(\cdot)}$ represents the conjugate operator. We can refer to \cite{Sun2018Inversion} for initialization and termination strategies.
    \begin{figure}[!t]
        \centering
        \subfloat[]{\includegraphics[height=0.34\linewidth]{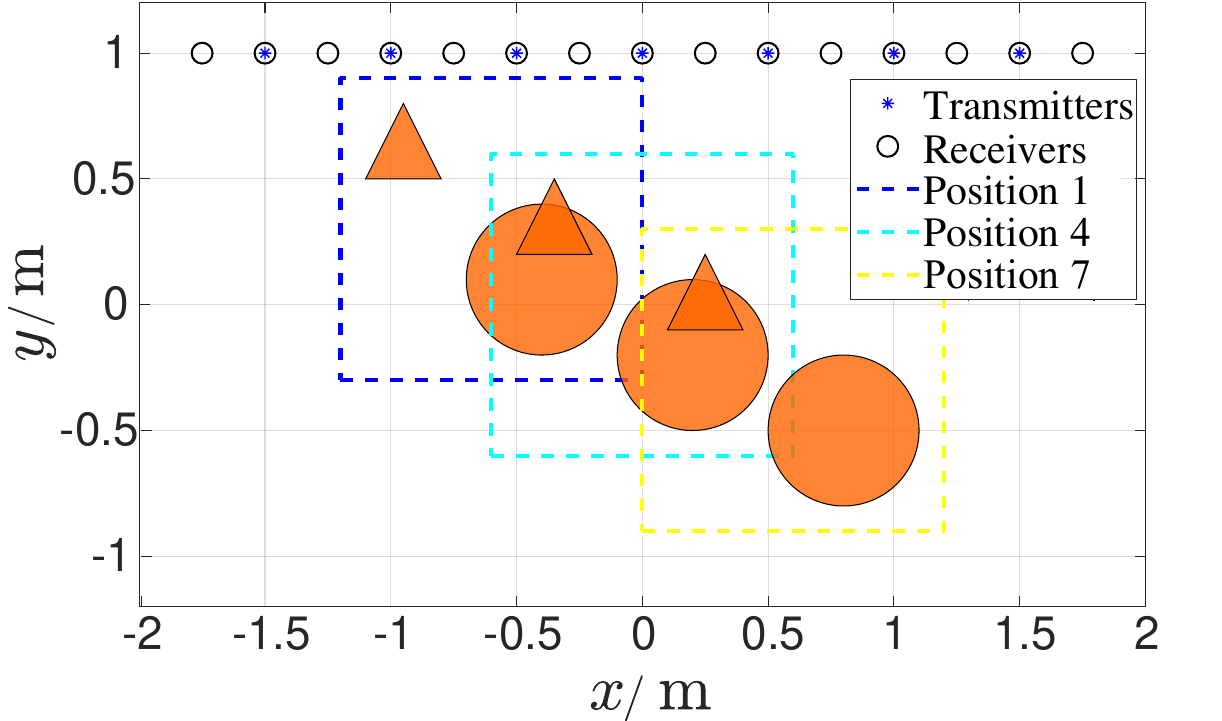}\label{fig:MeasureConfig}} 
        \subfloat[]{\includegraphics[height=0.3\linewidth]{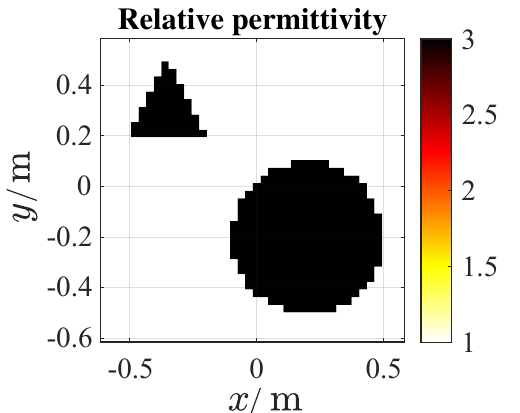}\label{fig:TriDisc3_GT}}%
        \caption{(a) The geometry of the \textit{TriDisc} profile and its synthetic measurement configuration. ~(b) Ground truth of \textit{TriDisc} profile at $m = 4$.}
    \end{figure}
    \begin{figure}[!t]
        \centering
        \subfloat[]{\includegraphics[width=0.33\linewidth]{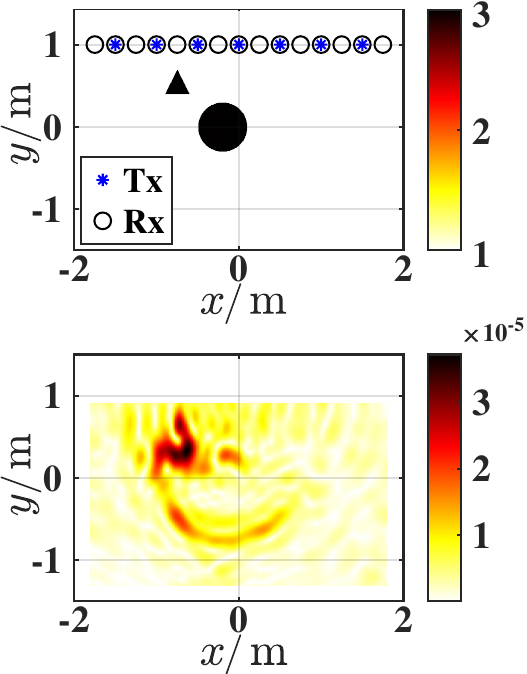}\label{fig:BPcycle1}}%
        \hfill
        \subfloat[]{\includegraphics[width=0.33\linewidth]{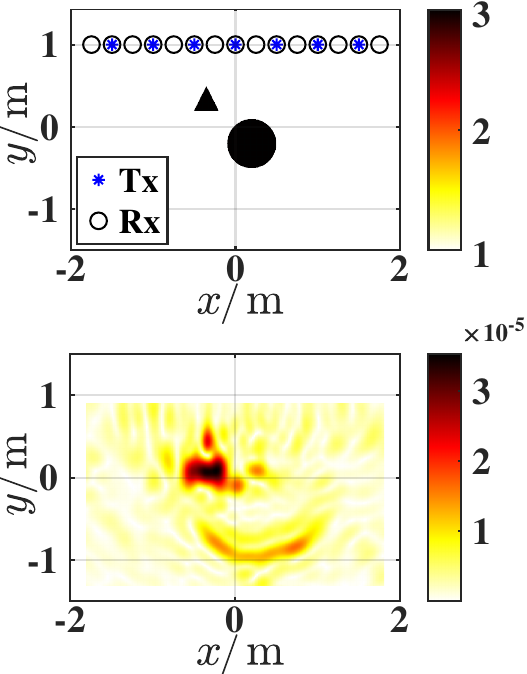}\label{fig:BPcycle3}}%
        \hfill 
        \subfloat[]{\includegraphics[width=0.33\linewidth]{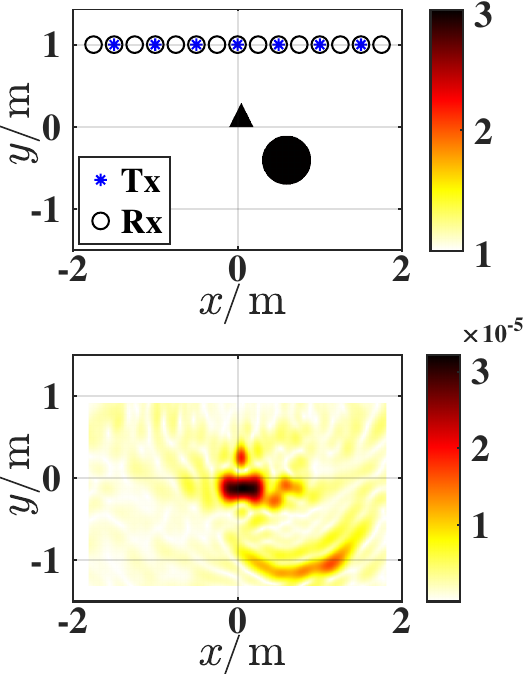}\label{fig:BPcycle5}}%
        \caption{The radar BP imaging results of \textit{TriDisc} at certain cycles of MIMO observation scanning. (a) Cycle 2. (b) Cycle 4. (c) Cycle 6. ~~(Top) Configuration and the ground truths $\varepsilon_r$. (Bottom) BP images [Amplitude].}
        \label{fig:1_5MIMOBP}   
    \end{figure}
    \begin{table}[!t]
        \renewcommand{\arraystretch}{1}
        \caption{Estimated relative center
coordinates of TriDisc dataset}
        \label{tab.BP_Rela_pos}
        \centering 
        \resizebox{\linewidth}{!}{
        \begin{tabular}{>{\centering\arraybackslash}m{1.25cm}|>{\centering\arraybackslash}m{6.7cm}}
        {SNR~[dB]}  & {${\hat{\bm{x}}}_{r,4}$(Upper)~[m],~${\hat{\bm{y}}}_{r,4}$(Lower)~[m]} \\ 
        \hline
        \multirow{2}*{$\infty$} & [-0.618, -0.407, -0.211, 0, 0.196, 0.392, 0.603] \\
        \cline{2-2}
           &  [0.321, 0.210, 0.111, 0, -0.100, -0.188, -0.287]  \\
        \cline{1-2}
        \multirow{2}*{10}  &  [-0.618, -0.407, -0.211, 0, 0.196, 0.392, 0.588]  \\ 
        \cline{2-2}
             &  [0.321, 0.210, 0.111, 0, -0.088, -0.188, -0.287]   \\
        \hline
        \end{tabular}}
    \end{table}
\section{Reconstruction Results}\label{sec.SimResults}
   This section introduces a 2-D benchmark profile \textit{TriDisc} to evaluate the performance of proposed method. Noise robustness is tested by adding complex Gaussian white noise to the total fields. Results are compared against static CC-CSI and ideal RMC-CC-CSI with known motion information (OPT-RMC-CC-CSI). The peak signal-to-noise ratio (PSNR) is used to assess the reconstruction accuracy. 2-D TM experimental data from an ultra-wideband system at National University of Defense Technology are used for further validation.
\subsection{Inversion of Synthetic Data}
Fig.~\ref{fig:MeasureConfig} shows the \textit{TriDisc} profile and measurement setup using 7 transmitters and 15 receivers. The relative permittivity is set to be $\varepsilon_r = 3$ with zero conductivity. The target moves with unknown velocity in trajectory from upper left to lower right over $N_{\text{M}}=7$ rounds of continuous observation. The ideal center positions at 7 observation instants are: ${\bm{x}}=[-0.6,-0.4,-0.2,0,0.2,0.4,0.6]$ m, ${\bm{y}}=[0.3,0.2,0.1,0,-0.1,-0.2,-0.3]$. The ground truth at $m = 4$ is shown in Fig.~\ref{fig:TriDisc3_GT}.
We choose all 7 rounds of MIMO data and arbitrarily select $c = 4$ as the reference positional center. This specific choice of $c$ solely enables controlled comparison for RMC-CC-CSI with static CC-CSI result at the same spatial position and does not affect the performance of RMC-CC-CSI. For $c=4$, $(x_4,y_4)=(0,0)$, according to~\eqref{eq.RelaPos} and $({\bm{x}},{\bm{y}})$, ideal relative coordinates $(\bm{x}_{r,4},\bm{y}_{r,4}) = ({\bm{x}},{\bm{y}})$. The BP imaging results at certain instants under frequencies spanning 0.1:0.05:1 GHz are shown in Fig.~\ref{fig:1_5MIMOBP} and $({\hat{\bm{x}}}_{r,4},{\hat{\bm{y}}}_{r,4})$ obtained through BP images registration under different SNR levels are shown in Table~\ref{tab.BP_Rela_pos}, which is close to ideal $({{\bm{x}}}_{r,4},{{\bm{y}}}_{r,4})$. Based on the BP imaging result at $c = 4$, domain $\mathcal{D}^4$ was established, employing frequencies of 0.3:0.1:1 GHz for inversion.

\begin{figure}[!t]
        \centering
        \subfloat[]{
            \begin{minipage}[t]{0.27\linewidth}
                \centering
                \includegraphics[width=\linewidth]{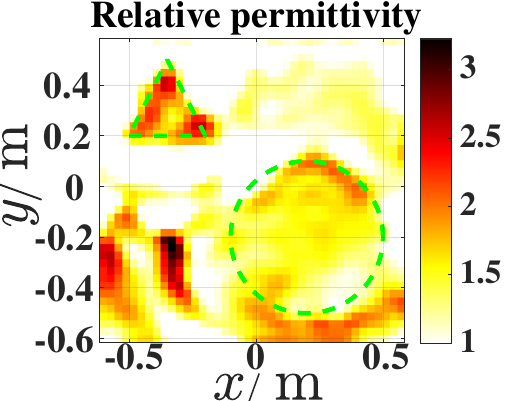}\label{fig:diel3_noisefree_it10000_inv_cccsi} \\
        \vspace{0.05cm}
                \includegraphics[width=\linewidth]{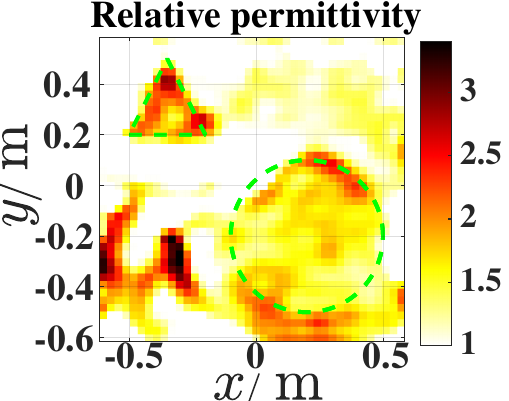}\label{fig:diel3_15dB_it10000_inv_cccsi} \\
        \vspace{0.05cm}
                \includegraphics[width=\linewidth]{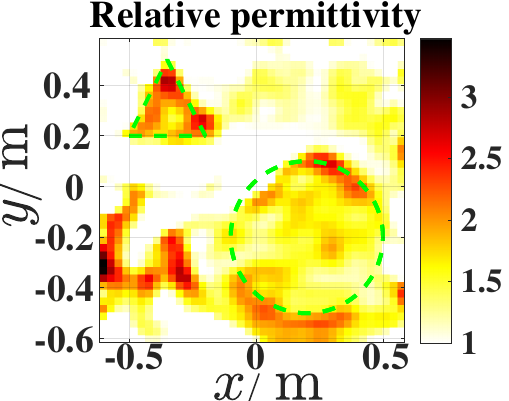}\label{fig:diel3_10dB_it10000_inv_cccsi}
            \end{minipage}\label{fig:diel3_cccsi} 
        }
        \hspace{0.3cm}
        \subfloat[]{
        \begin{minipage}[t]{0.27\linewidth}
            \centering
            \includegraphics[width=\linewidth]{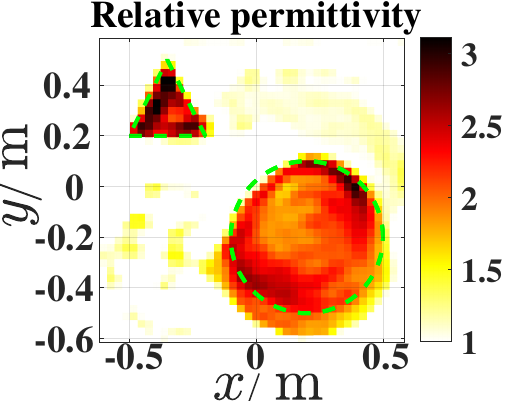}\label{fig:diel3_noisefree_it10000_inv_opt-tmt-cccsi} \\
                    \vspace{0.05cm}
            \includegraphics[width=\linewidth]{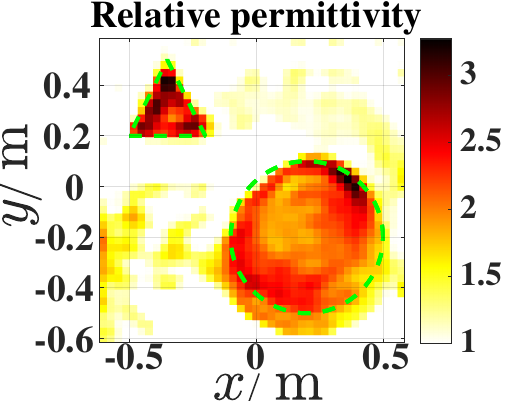}\label{fig:diel3_15dB_it10000_inv_opt-tmt-cccsi} \\
                    \vspace{0.05cm}
            \includegraphics[width=\linewidth]{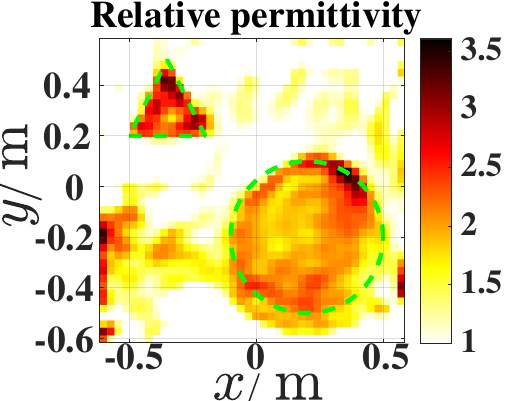}\label{fig:diel3_10dB_it10000_inv_opt-tmt-cccsi}
        \end{minipage}\label{fig:diel3_opt-tmt-cccsi}
        }
        \hspace{0.3cm}
        \subfloat[]{
        \begin{minipage}[t]{0.27\linewidth}
            \centering
            \includegraphics[width=\linewidth]{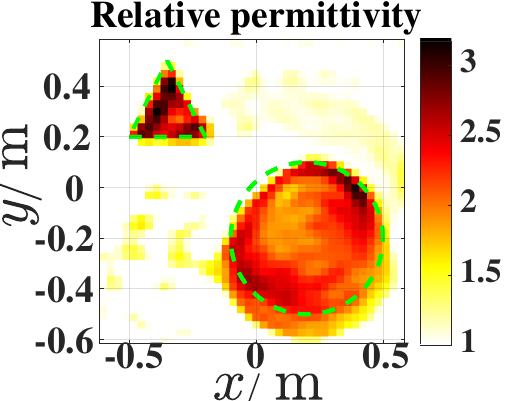}\label{fig:diel3_noisefree_it10000_inv_tmt-cccsi} \\
                    \vspace{0.05cm}
            \includegraphics[width=\linewidth]{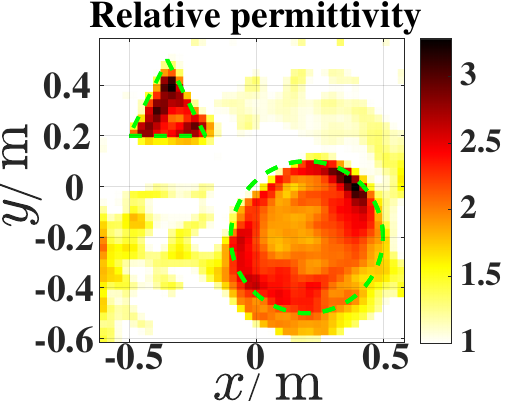}\label{fig:diel3_15dB_it10000_inv_tmt-cccsi} \\
                    \vspace{0.05cm}
            \includegraphics[width=\linewidth]{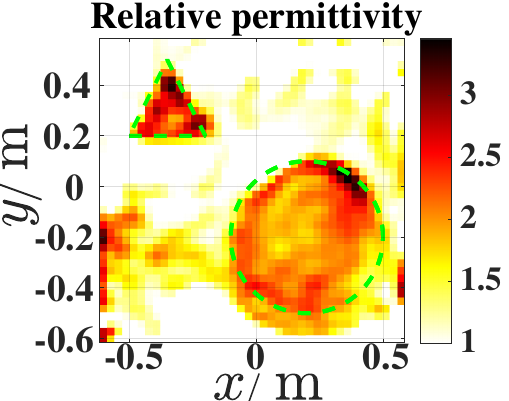}\label{fig:diel3_10dB_it10000_inv_tmt-cccsi}
        \end{minipage}\label{fig:diel3_tmt-cccsi}
        }
        \caption{The reconstructed optimal contrast profiles of \textit{TriDisc} obtained by different methods when $c$ = 4 during $ 5 \times 10^3$ iterations. The frequency range is 0.3:0.1:1 GHz. (Top) Noise-free. (Middle) SNR = 15 dB. (Bottom) SNR = 10 dB. (a) CC-CSI. (b) OPT-RMC-CC-CSI. (c) RMC-CC-CSI.}
        \label{fig:3Inv}   
    \end{figure}

\begin{figure}[!t]
        \centering
        \includegraphics[width=0.66\linewidth]{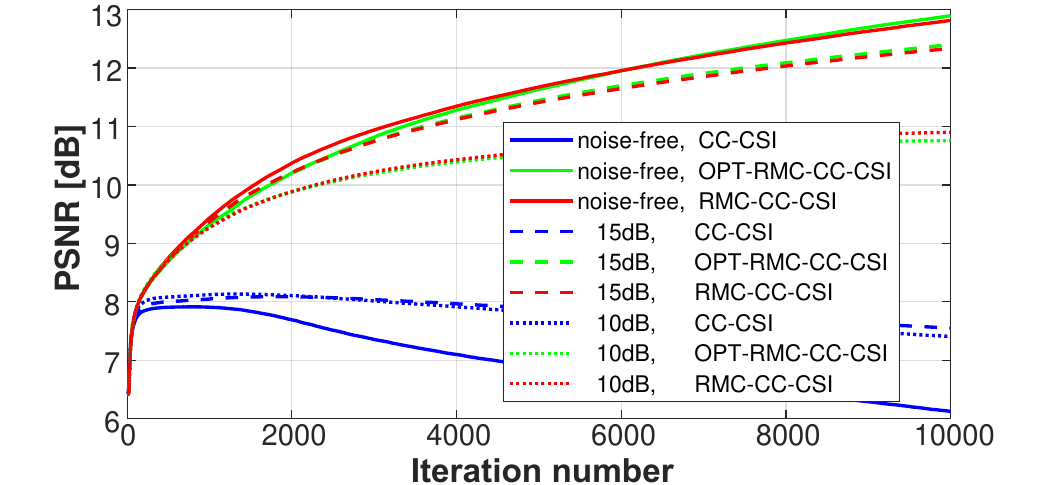}%
        \caption{PSNR curves in terms of the iteration number when processing the dataset \textit{TriDisc} by three inversion methods under different SNR levels.}
        \label{fig:3psnr}   
    \end{figure}

The results are shown in Fig.~\ref{fig:diel3_cccsi}-\ref{fig:diel3_tmt-cccsi}. Conventional CC-CSI converges to a local optimum under noise-free or noise interference conditions, whereas both OPT-RMC-CC-CSI and RMC-CC-CSI achieve more precise reconstruction. PSNR curves in Fig.~\ref{fig:3psnr} confirm that two RMC-type methods outperform CC-CSI in convergence speed, reconstruction accuracy, and noise immunity. The comparable performance between RMC-CC-CSI and OPT-RMC-CC-CSI also validates the efficacy of the BP image registration-based positioning method.

\subsection{Inversion of Experimental Data}
\begin{figure}[!t]
        \centering
        \subfloat[]{\includegraphics[width=0.4\linewidth]{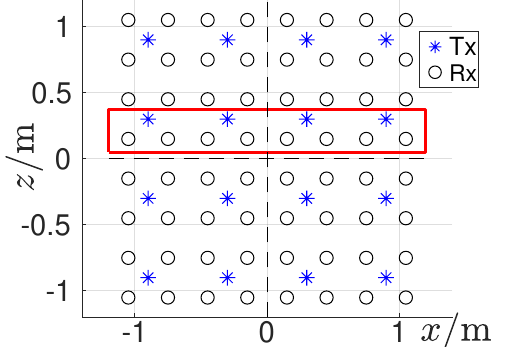}\label{fig:EXPconfig}} 
            \hfill           
        \subfloat[]{\includegraphics[height=0.23\linewidth]{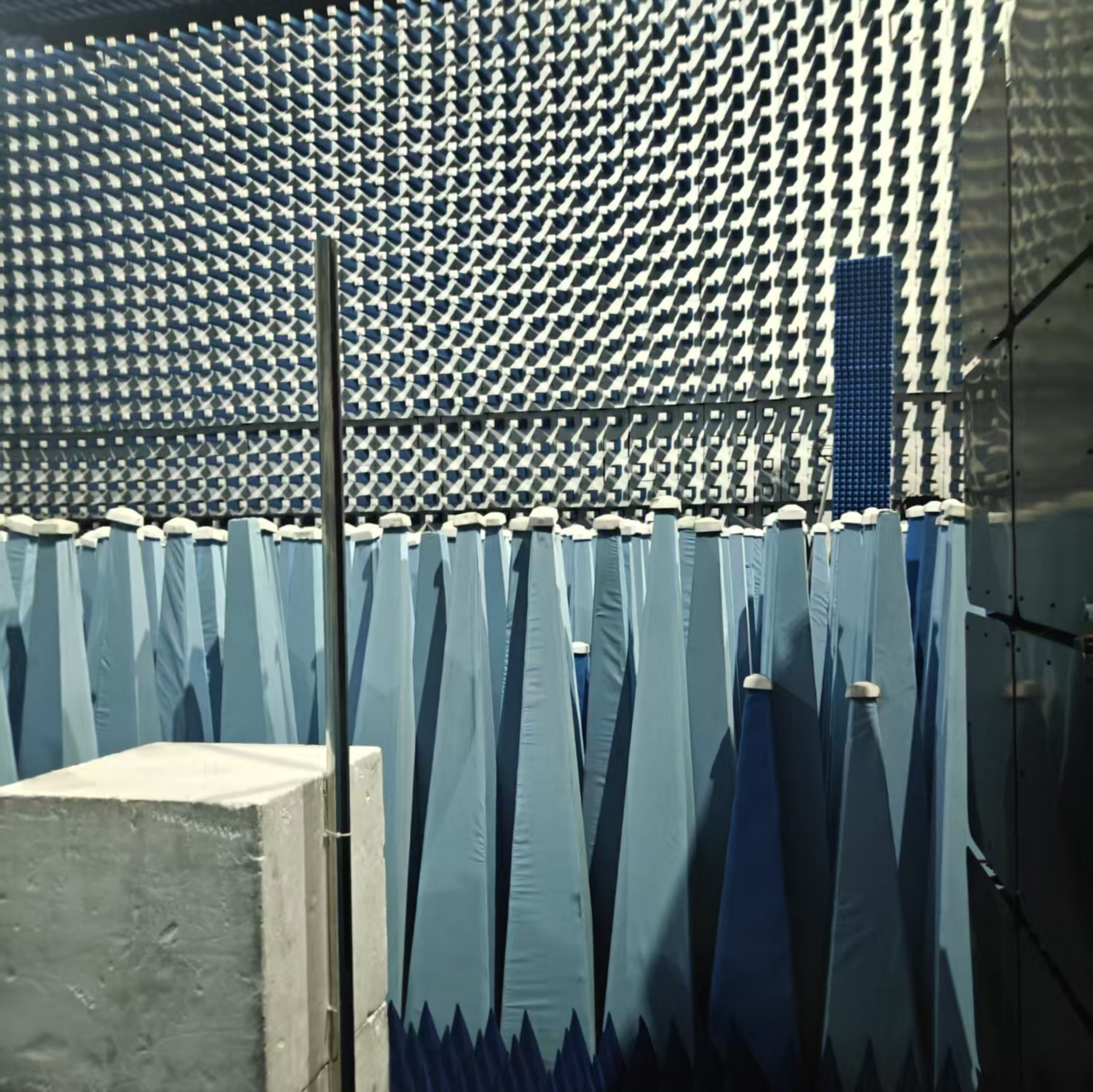}\label{fig:PECCylinder}} 
            \hfill
            \hspace{0.10cm}
        \subfloat[]{\includegraphics[height=0.23\linewidth]{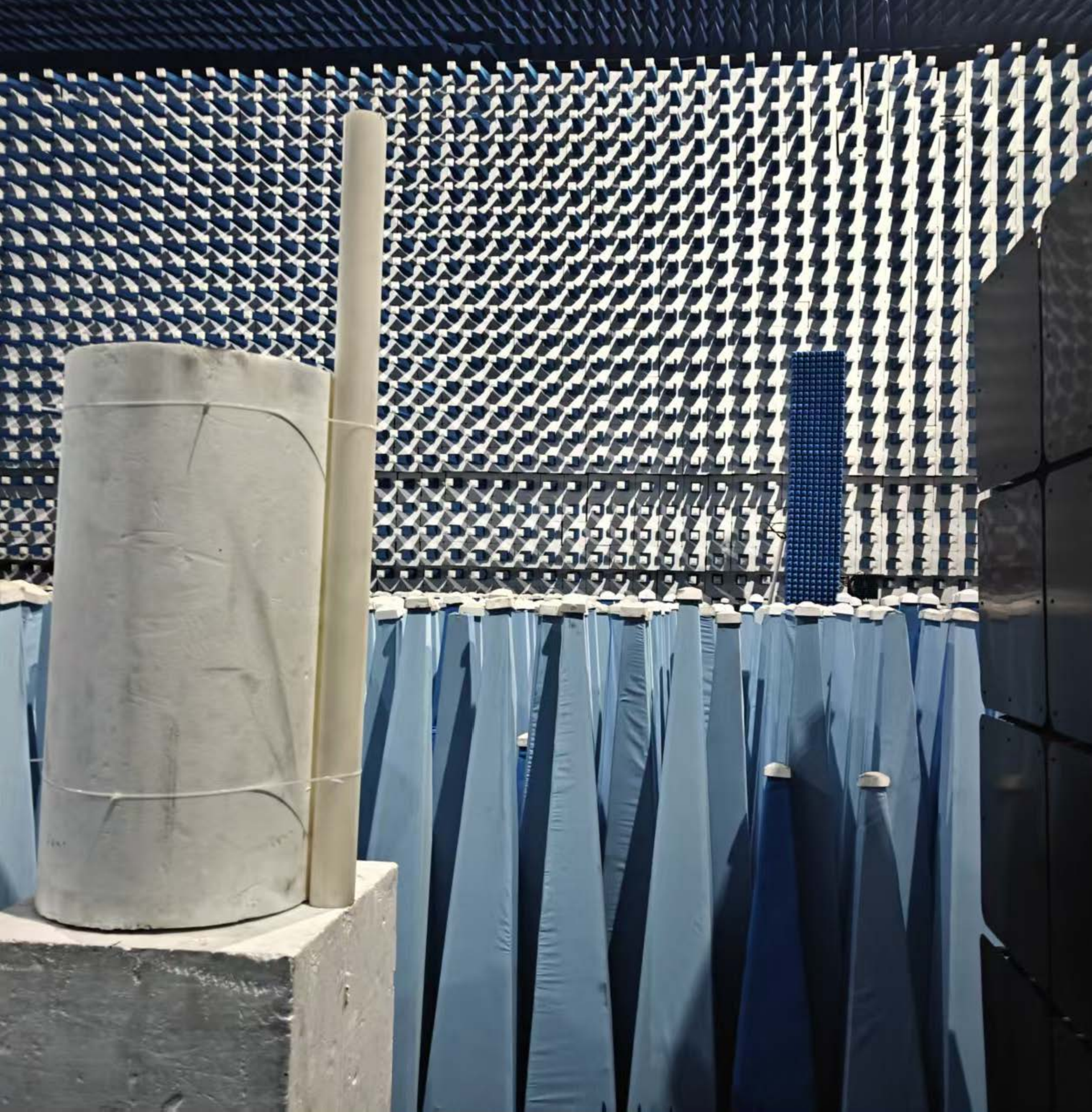}\label{fig:PA66Cylinder}}
        \caption{(a) MIMO array. (b) \textit{PEC\_Cylinder\_exp}. (c) \textit{PA66\_Cylinder\_exp}}
        \label{fig:array_datasets}
    \end{figure}

The ultra-wideband system (check more details in \cite{10418198}) employs transmitter and receiver spacings of 0.6 m and 0.3 m, with a 0.15 m height difference shown in Fig.~\ref{fig:EXPconfig}. We selected two adjacent rows of vertical transmitters and receivers marked in red box and maintained an adequate target-array spacing so that we can use cylinder with sufficient length to approximate a 2‑D TM measurement setup \cite{Jean}. Measurements followed a stop‑go‑stop protocol. A complex calibration procedure \cite{Geffrin} was applied before quantitative inversion. 

\begin{figure}[!t]
        \centering
        \subfloat[]{
        \begin{minipage}[t]{0.49\linewidth}
            \centering
            \includegraphics[width=\linewidth]{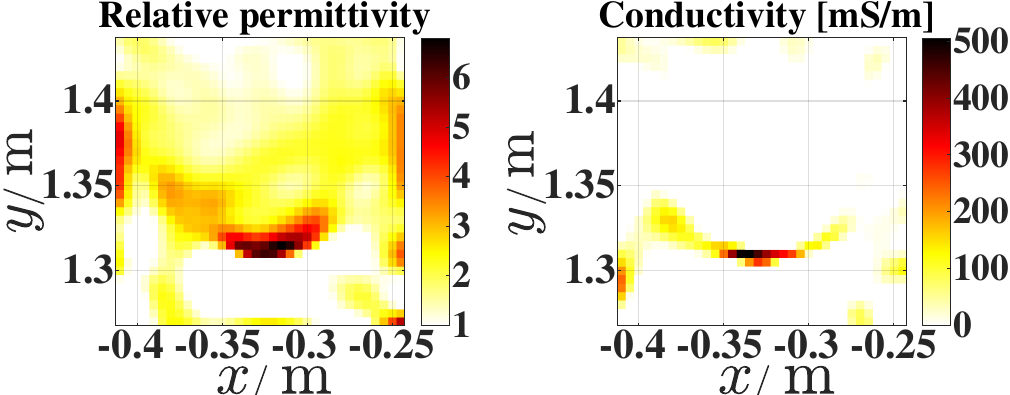}\label{fig:PECCylinder_cccsi} \\
            \vspace{0.11cm}
            \includegraphics[width=\linewidth]{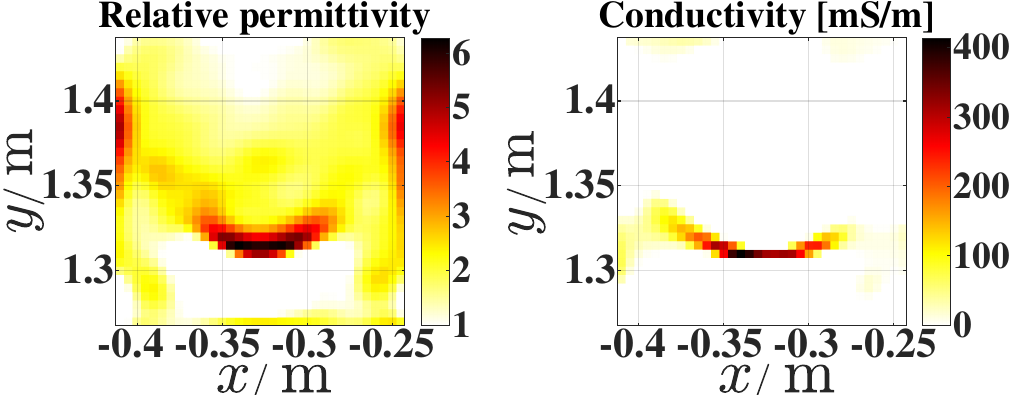}\label{fig:PECCylinder_rmc-cccsi}
        \end{minipage}\label{fig:PECCylinder_results}}
        \hfill
        \subfloat[]{
        \begin{minipage}[t]{0.48\linewidth}
            \centering
            \includegraphics[width=\linewidth]{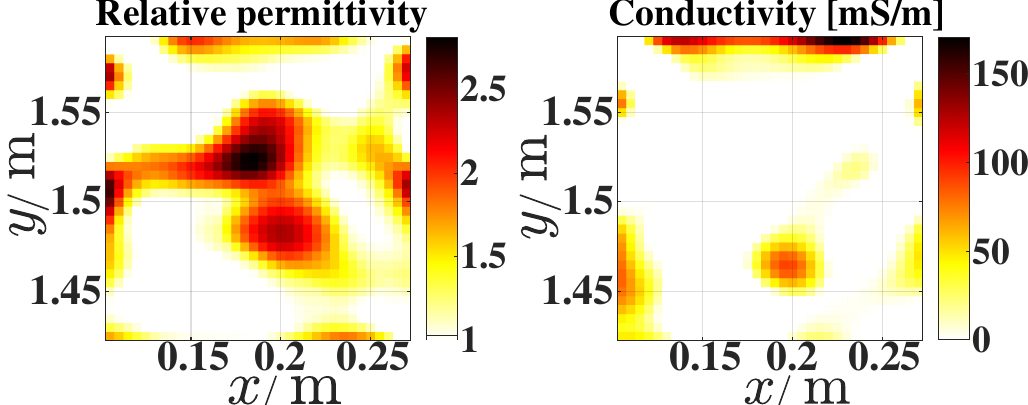}\label{fig:PA66_cccsi} \\ 
            \vspace{0.10cm}
            \includegraphics[width=\linewidth]{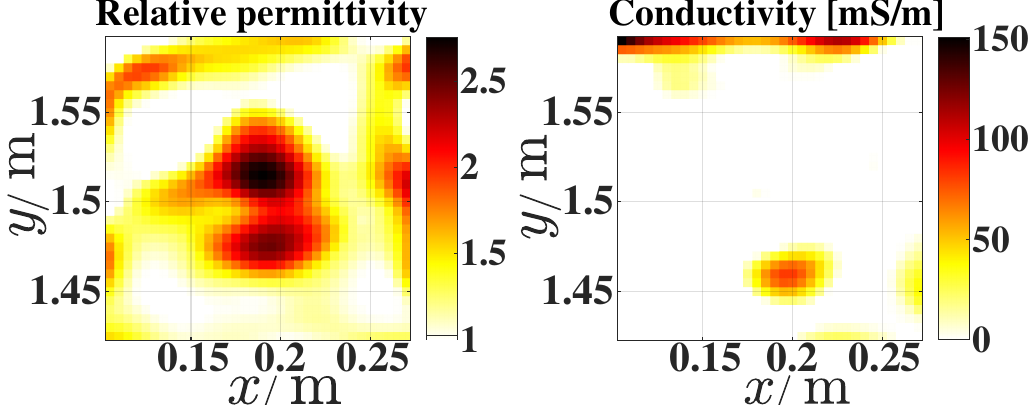}\label{fig:PA66_rmc-cccsi}
        \end{minipage}\label{fig:PA66Cylinder_results}}
        \caption{Experimental reconstructed profiles when $c$ = 1 during $1\times10^3$ iterations. (Top) CC-CSI. (Bottom) RMC-CC-CSI. (Left) $\varepsilon_r$. (Right) $\sigma$ [mS/m]. (a) \textit{PEC\_Cylinder\_exp}. (b) \textit{PA66\_Cylinder\_exp}.}
        \label{fig:Exp_Results}   
    \end{figure}

The \textit{PEC\_Cylinder\_exp} dataset was acquired by translating a radius of 25 mm PEC cylinder in Fig.~\ref{fig:PECCylinder} across $N_\text{M}=7$ spatial positions which moved from left to right parallel to the array. By employing BP imaging on all rounds of MIMO data, the estimated relative spatial positions $({\hat{\bm{x}}}_{r,1},{\hat{\bm{y}}}_{r,1})$ of \textit{PEC\_Cylinder\_exp} are calculated based on BP images registration with reference center $c$ = 1. Then, the data of 17 frequencies selected by equal wavelength between 1.5 GHz and 3.75 GHz are inverted. Similarly, data from translating a radius of 40 mm PA66 cylinder with $\varepsilon_r = 2.9\pm0.1$ to 5 different spatial positions ($N_{\text{M}} = 5$) shown in Fig.~\ref{fig:PA66Cylinder} is also processed. Reconstructed profiles by CC-CSI and RMC-CC-CSI are shown in Fig.~\ref{fig:Exp_Results}, where RMC-CC-CSI results show obviously better symmetry and align more closely with reality compared to that of the CC-CSI algorithm, validating the effectiveness of the proposed scheme. 

\section{Conclusion}
    In this paper, a new scheme for EM quantitative inversion of non-cooperative translationally moving targets is proposed, which achieves accurate localization and registration of translating targets before conducting the RMC-CC-CSI algorithm. Synthetic dataset demonstrates accelerated convergence, enhanced reconstruction accuracy and improved noise immunity. Experimental results validate its effectiveness. Future work will focus on solving ISPs for higher-contrast moving targets with lighter computational burdens. 
\bibliographystyle{IEEEtran}
\bibliography{mybib}

\newpage

\vfill

\end{document}